\documentclass[useAMS,usenatbib]{mn2e}

\usepackage{graphicx}% Include figure files
\usepackage{color}% color
\usepackage{amsmath, amssymb}
\usepackage{natbib}

\usepackage{ulem}

\title[Strong Coupling in RMHD Turbulence]
{Strong Coupling of Alfv\'en and Fast Modes in Compressible Relativistic Magnetohydrodynamic Turbulence in Magnetically-Dominated Plasmas}

\author[Takamoto, \& Lazarian]{M. Takamoto,$^{1}$\thanks{E-mail:
mtakamoto@eps.s.u-tokyo.ac.jp}
and
A. Lazarian$^2$\thanks{E-mail:
alazarian@facstaff.wisc.edu}
\\
$^{1}$Department of Earth and Planetary Science, University of Tokyo, Tokyo 113-0033, Japan \\
$^{2}${Department of Astronomy, University of Wisconsin, 475 North Charter Street, Madison, WI 53706, USA}
}
\begin{document}

\date{}

\pagerange{\pageref{firstpage}--\pageref{lastpage}} \pubyear{2017}

\maketitle

\label{firstpage}

\begin{abstract}
%\textcolor{red}{%
In this paper, 
we report our detailed analysis of the new strong-coupling regime between Alfv\'en and fast modes in Poynting-dominated plasma turbulence, 
reported in our previous work \citet{2016ApJ...831L..11T}, 
which is an important effect for many relativistic plasma phenomena,  
and calls for new theories of Poynting-dominated MHD turbulence. 
We performed numerical simulations of relativistic MHD turbulence in isothermal plasmas, 
and analyzed the ratio of fast to Alfv\'en mode energy. 
We found that 
the increase of the fast mode with the background $\sigma$-parameter can be observed even in isothermal plasma, 
showing that such a phenomena is universal in trans-Alfv\'enic turbulence in Poynting-dominated plasmas. 
To study the detailed energy conversion process, 
we also performed a series of simulations of decaying turbulence 
injecting pure Alfv\'en, fast, and slow modes, respectively, 
and investigated the development of the mode conversion from the each mode. 
We also found that 
the mode conversion between Alfv\'en and fast modes is nearly insensitive to the background temperature. 
Finally, 
we report a result of a simulation with initially fast mode dominated turbulence. 
It developed into a temporally strong-coupling regime, 
which is a strong evidence for the existence of our suggesting strong-coupling regime of fast and Alfv\'en modes. 
Our result suggests that 
the strong turbulence in Poynting-dominated plasma is very different from that in the 
non-relativistic plasma. 
It will also give an important guidance to studies of particle acceleration and non-thermal photon emission 
from Poynting-dominated plasma. 
%}%
\end{abstract}


\section{Introduction}
\label{sec:sec1}

Turbulence plays an important role in many astrophysical phenomena. 
In particular, 
the effect of magnetic field is essential in many cases, 
and there has been a lot of studies on magnetohydrodynamic (MHD) turbulence. 
In non-relativistic case, 
it has been recognized that 
the coupling of each MHD characteristic mode is very weak, 
and they can be treated separately \citep{2002PhRvL..88x5001C,2003MNRAS.345..325C}, 
which provided with a vast amount of applicability of the critical-balance turbulence \citep{1995ApJ...438..763G} 
and Kolmogorov turbulence \citep{1941DoSSR..30..301K} to various kinds of astrophysical phenomena. 
Comparing with non-relativistic work, 
much less attention has been given to relativistic turbulence 
\citep{1998PhRvD..57.3219T,2005ApJ...621..324C,2011ApJ...734...77I,2012ApJ...744...32Z,2013ApJ...763L..12Z,2013ApJ...766L..10R,2014ApJ...780...30C,2015ApJ...815...16T}. 
Recently, \citet{2016ApJ...831L..11T} (TL16 in the following) has performed a mode decomposition of relativistic MHD turbulence, 
and reported that 
the coupling of each mode in relativistic MHD turbulence became stronger with increasing the background relativistic magnetization parameter $\sigma$, 
where $\sigma$ is defined as $\sigma \equiv B_0^2/4 \pi \rho h c^2 \gamma^2$ 
where $B_0$ is the background magnetic field, $\rho$ is the rest mass density, $h$ is the specific enthalpy, 
$c$ is the velocity of light, and $\gamma$ is the Lorentz factor. 
However, 
TL16 assumed an adiabatic plasma 
which allowed increasing of the background temperature through thermalization of turbulence kinetic energy. 
In addition, 
it did not clarify the properties of the turbulence in the strong-coupling regime. 

In this paper, 
we report our detailed analysis of the new strong-coupling regime between Alfv\'en and fast modes in Poynting-dominated plasma turbulence, 
reported in our previous work TL16. 
We performed numerical simulations of relativistic MHD turbulence in isothermal plasmas, 
and analyzed the ratio of fast to Alfv\'en mode energy. 
We also performed a simulation with initially fast mode dominated turbulence 
which allowed us to obtain temporally strong-coupling regime. 

\section{Numerical Setup}
\label{sec:sec2}

The plasma is modeled by the ideal RMHD approximation with the TM equation of state \citep{2005ApJS..160..199M} 
that allows us to simulate the relativistic perfect gas equation of state \citep{synge1957relativistic} with less than 4 \% error. 
%\textcolor{blue}{%
The equations are updated using a numerical code originally developed by \citep{2011ApJ...734...77I}
that combines the relativistic HLLD method \citep{2009MNRAS.393.1141M} in a conservative fashion 
and the constrained transport algorithm \citep{1988ApJ...332..659E,2005JCoPh.205..509G}. 
%}%
The initial background plasma is assumed to be uniform with magnetic field ${\bf B}_0$, density $\rho_0$, and temperature $k_{\rm B} T_0/m c^2$ 
where $k_B$ is the Boltzmann constant, $T_0$ is the temperature, $m$ is the particle mass. 
In the following, we set $k_{\rm B} = 1$ for simplicity. 
In our previous work TL16, 
a simple ideal RMHD plasma was considered 
which allowed its temperature variation from injected turbulence. 
Although we checked that our results were insensitive to the background temperature, 
it may have caused a suspicion that our results, indicating fast mode increasing in Poynting-dominated plasmas, 
was due to the background temperature increase. 
For this reason, 
an isothermal plasma is assumed in this work, 
that is, we solved the ideal RMHD equation every time step, 
%and set temperature to the ititial value at every 0.1 eddy-turnover time. 
and introduced an cooling in temperature %, not the total energy, 
whose timescale was 0.1 eddy-turnover time 
\footnote{
%\textcolor{red}{%
In our work, we did not use isothermal equation of state 
because of the numerical stability for solving trans-Alfv\'enic turbulence 
in high-$\sigma$ plasma which is usually very difficult to solve 
even when satisfying the conservation of energy. 
Note that this procedure in principle allows heating of the plasma for some regions, 
for example, expanding regions. 
It is, however, very rare because of the uniform injection of strong turbulence. 
A similar strategy was taken by \citet{2013ApJ...763L..12Z}. 
%}%
}
.

In our work, 
the amount of Poynting energy in the plasma is measured by the following parameter $\sigma$, which is defined as
\begin{equation}
  \sigma \equiv \frac{B_0^2}{4 \pi \rho_0 h_0 c^2 \gamma_0^2}
  ,
  \label{eq:2.1}
\end{equation}
where $h$ is the specific enthalpy determined by the TM equation of state, 
and $\gamma$ is the Lorentz factor. 
The subscript 0 means the value of the background plasma. 
Originally, this parameter was defined by the ratio of the Poynting energy flux to particle energy flux; 
it reduces to the above form when considering the MHD state (${\bf E} = - {\bf v} \times {\bf B}$) with flowing perpendicular to the background magnetic field. 
Note that, in the case of RMHD turbulence, it is expected that 
the basic properties can be governed by $\sigma$-parameter and the temperature $T$ 
because the RMHD characteristic velocities can be described by those 2-parameters and the angle between the propagation direction and the magnetic field 
(see also Equation (\ref{eq:3.23})). 

An isotropic turbulent velocity is injected at the initial time-step, simulating so-called \textit{decaying turbulence}. 
Similar to \citep{2015ApJ...815...16T} and TL16, 
the turbulence is injected at large scales, $l_{\rm inj} = L/2, L/3, L/4$, where $L$ is the numerical box size, 
whose energy spectrum is assumed to be flat. 
%Following \citep{2002PhRvL..88x5001C,2003MNRAS.345..325C}, 
%The turbulence is injected only with an Alfv\'en mode velocity component. 
%that is obtained by the method explained in the next section. 
We consider a cubic numerical domain that is divided by uniform meshes 
whose size is typically $\Delta = L/512$; 
the higher resolutions, $\Delta = L/1024$, is used for obtaining the strong-coupling regime 
provided in Section \ref{sec:sec4.3}. 
%The resolution is chosen as sufficient for resolving turbulent-eddies 
%responsible for the mode exchange in our simulation.

\section{Mode Decomposition of RMHD Turbulence}
\label{sec:sec3}

\subsection{Derivation of the Relativistic Displacement Vectors of RMHD Modes}
\label{sec:sec3.1}

There have already been a lot of references of linear perturbation of ideal RMHD equations. 
In this section, we just briefly introduce our derivation of the mode decomposition of the RMHD characteristic modes. 
In the following, we consider the analysis only in the fluid rest frame 
which is also used for our numerical simulations. 
The linear perturbation in the general frame was provided, for example, in \citep{1990rfmf.book.....A,1999MNRAS.303..343K,2001ApJS..132...83B,2010ApJS..188....1A}. 
The ideal RMHD equations are given as: 
\begin{align}
  &\partial_t (\rho \gamma) + \nabla \cdot (\rho \gamma {\bf v}) = 0, 
  \label{eq:3.1}
  \\                                                                
  &\partial_t [(\rho h + b^2) u^t u_j - b^t b_j] 
  \nonumber
  \\
  &+ \nabla_i [ (\rho h + b^2) u^i u_j + (p_{\rm g} + b^2/2) \delta^i_j - b^i b_j] = 0, 
  \label{eq:3.2}
  \\                                                                
  &\partial_t {\bf B} = \nabla \times ({\bf v} \times {\bf B}), \quad \nabla \cdot {\bf B} = 0,
  \label{eq:3.3}
\end{align}
where $i, j$ run from 1 to 3 following the Einstein rule, 
$\delta^i_j$ is the identity matrix, 
$u^{\mu} = \gamma (1, {\bf v})$ is the four velocity, 
and $b^{\mu} = \gamma [ {\bf v \cdot B}, {\bf B}/\gamma^2 + ({\bf v \cdot B}){\bf v}]$ is the covariant magnetic field 
which is the magnetic field in the fluid comoving frame. 
Note that the magnetic field is redefined as to absorb the coefficients, such as $4 \pi$ in the case of the Gauss unit. 
The perturbed variables are assumed to be written as: $\delta Q \propto \exp[- i (\omega t + {\bf x \cdot k})]$, 
and the unperturbed variables are indicated by the subscript 0 as, $Q_{\rm 0}$.
We assume an adiabatic equation of state, $\delta p = c_{\rm s}^2 h \delta \rho$ 
where $c_{\rm s}$ is the sound velocity depending on the background temperature $T$ \citep{2005ApJS..160..199M}. 
The perturbed equations of the above ones are given as:
\begin{align}
  &- i \omega \delta \rho + i {\bf k} \cdot (\rho_0 \delta {\bf v}) = 0, 
  \label{eq:3.4}
  \\                                                                
  &- i \omega [(\rho_0 h_0 + B_0^2) \delta v^j - (\delta {\bf v} \cdot {\rm B_0}) B_0^j] 
  \nonumber
  \\
  &+ i k_i [ (c_{\rm s}^2 h_0 \delta \rho + {\bf B}_0 \cdot \delta {\bf B}) \delta^{ij} - B_0^j \delta B^i - B_0^i \delta B^j] = 0, 
  \label{eq:3.5}
  \\                                                                
  &- i \omega \delta {\bf B} = i {\bf k} \times (\delta {\bf v} \times {\bf B}_0), \quad i {\bf k} \cdot \delta {\bf B} = 0.
  \label{eq:3.6}
\end{align}
Similarly to the non-relativistic case, 
the above equations can be divided as:
\begin{align}
  &- \omega \delta \rho/\rho_0 + k_x \delta v_x + k_y \delta v_y = 0, 
  \label{eq:3.7}
  \\                                                                
  &- \omega \rho_0 h_0 \delta v^x + k_x (c_{\rm s}^2 h_0 \delta \rho) = 0, 
  \label{eq:3.8}
  \\
  &- \omega (\rho_0 h_0 + B_0^2) \delta v^y 
  + k_y (c_{\rm s}^2 h_0 \delta \rho + B_0 \delta B_x) - k_x B_0 \delta B^y = 0, 
  \label{eq:3.9}
  \\                                                                
  &- \omega \delta B^y = B_0 k_x \delta v^y, 
  \label{eq:3.10}
  \\
  &\delta B_x = - \frac{k_y \delta B_y}{k_x},
  \label{eq:3.11}
\end{align}
and 
\begin{align}
  &- \omega (\rho_0 h_0 + B_0^2) \delta v^z - k_x B_0 \delta B^z = 0, 
  \label{eq:3.12}
  \\
  &- \omega \delta B^z = B_0 k_x \delta v^z, 
  \label{eq:3.13}
\end{align}
where the magnetic field is set in the x-direction, and the wave vector ${\bf k}$ is assumed in the x-y plane. 
%Note that the degree of freedom of $\delta B_x$ is not considered here because of the divergence free condition Equation (\ref{eq:3.6}). 
It is well-known that the first set describes the fast and slow modes, and the second set describes the Alfv\'en mode. 
From the second set, it is clear that the Alfv\'en mode velocity can be obtained 
by projecting the velocity on the direction ${\bf k} \times {\bf B}_0$. 
The decomposition of fluid velocity onto fast and slow mode can be obtained 
%by \textcolor{red}{eliminating} $\delta \rho, \delta B_x, \delta B_y$ from Equations (\ref{eq:3.7}) to (\ref{eq:3.11}). 
by eliminating $\delta \rho, \delta B_x, \delta B_y$ from Equations (\ref{eq:3.7}) to (\ref{eq:3.11}). 
From Equations (\ref{eq:3.7}) and (\ref{eq:3.8}), 
we obtain 
\begin{equation}
  \hat{\xi} \propto k_x \hat{k}_x + \left[\frac{u^2}{c_{\rm s}^2} k^2 - k_x^2 \right] \frac{1}{k_y^2} k_y \hat{k}_y
  ,
  \label{eq:3.14}
\end{equation}
where $\hat{k}_i$ is a unit vector in the i-th direction. 
Similarly, using Equations (\ref{eq:3.7}), (\ref{eq:3.9}), (\ref{eq:3.10}) and (\ref{eq:3.11}), 
we obtain 
\begin{equation}
  \hat{\xi} \propto \left[u^2 (1 + \sigma) k^2 - c_{\rm s}^2 k_y^2 - \sigma k^2 \right] \frac{1}{c_{\rm s}^2 k_x^2} k_x \hat{k}_x + k_y \hat{k}_y.
  \label{eq:3.15}
\end{equation}
Substituting the slow velocity $u_{\rm slow}$ in Equation (\ref{eq:3.14}), and the fast velocity $u_{\rm fast}$ in Equation (\ref{eq:3.15}), 
which, in the fluid comoving frame, are given as:
\begin{align}
  u_{\rm fast/slow}^2 &= \frac{1}{2} \left[ 
   c_{\rm A}^2 + \left( \cos^2 \theta + \frac{\sin^2 \theta}{1 + \sigma} \right) c_{\rm s}^2
   \right.
   \nonumber                    
   \\
   &\left. \pm \sqrt{\left\{ c_{\rm A}^2 + \left( \cos^2 \theta + \frac{\sin^2 \theta}{1 + \sigma} \right) c_{\rm s}^2 \right\}^2 - 4 c_{\rm s}^2 c_{\rm A}^2 \cos^2 \theta}
   \right]
   ,
  \label{eq:3.23}  
\end{align}
and the Equations (4) and (5) in TL16 are reproduced, 
which are also used for mode decomposition in the following. 
 
\subsection{Non-Relativistic Limit}
\label{sec:sec3.2}

In the non-relativistic limit, 
Equations (\ref{eq:3.14}) and (\ref{eq:3.15}) reduces to the Equation (1) and (2) in \citep{2002PhRvL..88x5001C} 
which are given as:
\begin{align}
  \hat{\xi}_{\rm slow} &\propto k_x \hat{k}_x + \frac{1 - \sqrt{D} - \beta/2}{1 + \sqrt{D} + \beta/2} \left[ \frac{k_x}{k_y} \right]^2 k_y \hat{k}_y
  ,
  \label{eq:3.16}
  \\
  \hat{\xi}_{\rm fast} &\propto \frac{1 - \sqrt{D} + \beta/2}{1 + \sqrt{D} - \beta/2} \left[ \frac{k_y}{k_x} \right]^2 k_x \hat{k}_x + k_y \hat{k}_y
  ,
  \label{eq:3.17}
\end{align}
where $D \equiv (1 + \beta/2)^2 - 2 \beta \cos^2 \theta$ and $\theta$ is the angle between magnetic field and wave vector: ${\bf k \cdot B_0} = k B_0 \cos \theta = k_x B_0$. 
After some calculations, the above equations can be rewritten as 
\begin{align}
  \hat{\xi}_{\rm slow} &\propto k_x \hat{k}_x + \left(\frac{1 + \beta/2 - \sqrt{D}}{\beta \cos^2 \theta} - 1\right) \left[ \frac{k_x}{k_y} \right]^2 k_y \hat{k}_y
  ,
  \label{eq:3.18}
  \\
  \hat{\xi}_{\rm fast} &\propto \left(\frac{(1 + \beta/2 + \sqrt{D}) -2}{\beta \sin^2 \theta} - 1 \right) \left[ \frac{k_y}{k_x} \right]^2 k_x \hat{k}_x + k_y \hat{k}_y
  .
  \label{eq:3.19}
\end{align}
We start from the slow mode which can be obtained by substituting $u = u_{\rm slow}$ into Equation (\ref{eq:3.14}). 
It reduces to
\begin{equation}
  \hat{\xi}_{\rm slow} \propto k_x \hat{k}_x + \left[\frac{u_{\rm slow}^2}{c_{\rm s}^2 \cos^2 \theta} - 1 \right] \left[ \frac{k_x}{k_y} \right]^2 k_y \hat{k}_y
  . 
  \label{eq:3.20}
\end{equation}
\citet{2002PhRvL..88x5001C} considered an isothermal plasma, 
and the sound velocity can be rewritten as $c_{\rm s}^2 = c_{\rm A}^2 \beta /2$. 
In the non-relativistic case, 
the fast and slow velocity can be written as: 
\begin{equation}
  u_{\rm fast/slow}^2 = \frac{c_{\rm A}^2}{2} \left[1 + \frac{\beta}{2} \pm \sqrt{D} \right]
  ,
  \label{eq:3.21}  
\end{equation}
where $+$ gives the fast velocity, and $-$ the slow velocity. 
From these relations, it is clear that Equation (\ref{eq:3.20}) reduces to Equation (\ref{eq:3.18}). 
Next, in the non-relativistic limit, Equation (\ref{eq:3.15}) can be rewritten as 
\begin{equation}
  \hat{\xi}_{\rm fast} \propto \left[\frac{u_{\rm fast}^2 - c_{\rm A}^2}{c_{\rm s}^2 \sin^2 \theta} - 1 \right] \left[ \frac{k_y}{k_x} \right]^2 k_x \hat{k}_x + k_y \hat{k}_y
  ,
  \label{eq:3.22}
\end{equation}
where $u = u_{\rm fast}$ and $\sigma \simeq c_{\rm A}^2$ are substituted. 
Similarly, it is clear that Equation (\ref{eq:3.22}) is equivalent to Equation (\ref{eq:3.19}). 

\subsection{Relativistic Corrections}
\label{sec:sec3.3}

In the strong magnetic field limit: $\beta \rightarrow 0$, 
the non-relativistic limit expressions, Equations (\ref{eq:3.16}) and (\ref{eq:3.17}), show that 
the slow mode displacement vector becomes parallel to the magnetic field, 
and the fast mode displacement vector becomes perpendicular to the magnetic field. 
In the following, 
we derive the displacement vectors in the relativistic plasma. 

The corresponding limit in relativistic plasmas is the high-$\sigma$ limit, 
that is, the electromagnetic field energy is larger than the plasma rest mass energy. 
In the non-relativistic case, 
the sound velocity becomes negligibly small comparing with the Alfv\'en velocity in this limit. 
In the relativistic case, however, the sound velocity can be in general not so small comparing with the Alfv\'en velocity, $0 \ll c_{\rm s} < c_{\rm A} \lesssim c$, 
because the Alfv\'en velocity is bounded by the light velocity. 
This can occur when the background plasma temperature reaches around the rest mass energy, $T \lesssim m c^2$. 
Even 0.1\% of the rest mass energy can make the sound velocity around 4\% of the light velocity, 
which is considered to be realized in many high-energy astrophysical phenomena. 
In this limit, $\sigma \gg 1$ and $T / mc^2 \lesssim 1$, 
the slow and fast mode velocities become 
\begin{align}
  u_{\rm fast}^2 &\simeq c_{\rm A}^2 + \frac{c_{\rm A}^2 c_{\rm s}^2 \sin^2 \theta}{\sigma (c_{\rm A}^2 - c_{\rm s}^2 \cos^2 \theta)}
  ,
  \label{eq:3.24}
  \\
  u_{\rm slow}^2 &\simeq c_{\rm s}^2 \cos^2 \theta
  ,
  \label{eq:3.25}
\end{align}
Substituting Equation (\ref{eq:3.24}) into Equation (\ref{eq:3.15}), 
the displacement vector becomes 
\begin{equation}
  \hat{\xi}_{\rm fast} \propto \frac{c_{\rm s}^2 \cos^2 \theta}{c_{\rm A}^2 - c_{\rm s}^2 \cos^2 \theta} \left[ \frac{k_y}{k_x}\right]^2 k_x \hat{k}_x + k_y \hat{k}_y
  . 
  \label{eq:3.26}
\end{equation}
Hence, the fast mode displacement vector has non-zero components parallel to the background magnetic field even in the high-$\sigma$ limit. 
Note that here we consider non-zero temperature 
which is the crucial difference from the force-free plasma discussed in \citep{1998PhRvD..57.3219T}, 
and this makes our results more realistic 
because, in high-$\sigma$ plasmas, dissipation of small magnetic field fluctuations can result in thermal energy comparable to rest mass energy 
\citep{2014ApJ...787...84T}. 
Similarly, substituting Equation (\ref{eq:3.25}) into Equation (\ref{eq:3.14}), 
the slow mode displacement vector reduces to the same result as the non-relativistic case, 
that is, parallel to the background magnetic field direction. 
Note that in the high-$\sigma$ and low-temperature limit, 
Equations (\ref{eq:3.24}) and (\ref{eq:3.25}) reduce to 
$u_{\rm fast}^2 \simeq c_{\rm A}^2 + c_{\rm s}^2 \sin^2 \theta /(1 + \sigma)$, and $u_{\rm slow} \simeq c_{\rm s}^2 \cos^2 \theta$, 
which give the same result as the non-relativistic case. 

\section{Mode Coupling Study}
\label{sec:sec4}

In this section, 
we discuss the mode coupling between fast, slow, and Alfv\'en modes, 
and their dependence on the background $\sigma$-parameter. 

\subsection{Ratio Between Compressible to Alfv\'en Modes}
\label{sec:sec4.1}

\begin{figure}%[t]
 \centering
  \includegraphics[width=8.cm,clip]{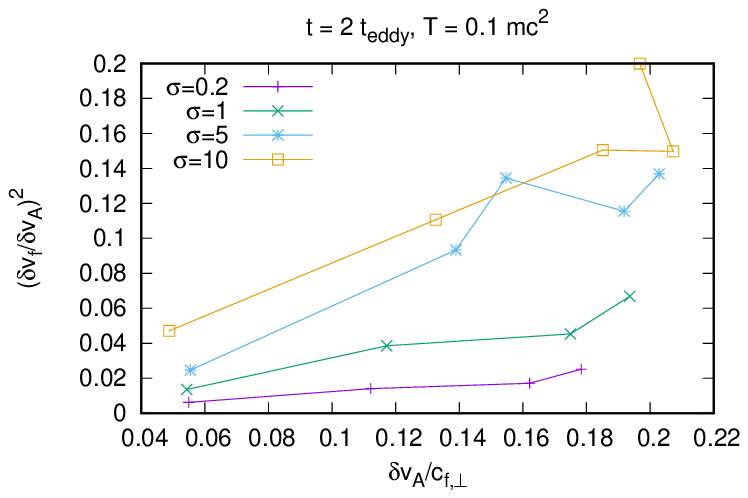}
  \includegraphics[width=8.cm,clip]{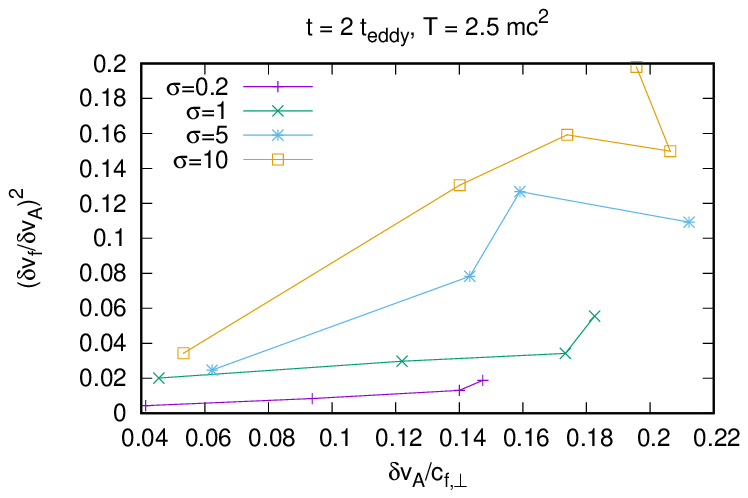}
  \caption{The ratio of fast to Alfv\'en mode power 
           in terms of the non-relativistic fast Mach number 
           %\textcolor{red}{%
           which is measured using the Alfv\'en mode velocity 
           at two eddy-turnover times. 
           %}%
           Top: $T=0.1 mc^2$, Bottom: $T=2.5 mc^2$.
           }
  \label{fig:1.1}
\end{figure}

We performed numerical simulations of decaying turbulence. 
Similar to our previous work TL16, 
the Alfv\'enic mode velocity was injected initially, 
and we simulated the temporal evolution of the mode exchange in an isothermal plasma. 

Figure \ref{fig:1.1} is the fast to Alfv\'en mode ratio at two eddy-turnover times 
where the eddy-turnover time is defined as $t_{\rm eddy} \equiv L/v_{\rm inj}$. 
The top panel is the results of cold plasma case with $T / mc^2 = 0.1$, 
and bottom panel is the results of hot plasma case with $T / mc^2 = 2.5$. 
%The horizontal axis is the injection velocity in the unit of the fast velocity 
%\textcolor{red}{%
Following \citep{2002PhRvL..88x5001C,2002ApJ...564..291C},
the horizontal axis is the Alfv\'en mode velocity at two eddy-turnover times 
%}%
in the unit of the fast velocity 
in the direction perpendicular to the back ground magnetic field
\footnote{
Note that Equation (\ref{eq:3.23}) shows that the fast velocity has its maximum at the perpendicular 
direction to the background magnetic field. 
We use this maximum value for the unit of the injection velocity, 
which provides with a better correlation to the ratio of fast to Alfv\'en power. 
}. 
Both show an increase of the ratio with the Alfv\'en mode velocity, 
and also with the background $\sigma$ value, 
which is consistent with TL16. 

\begin{figure}%[t]
 \centering
  \includegraphics[width=8.cm,clip]{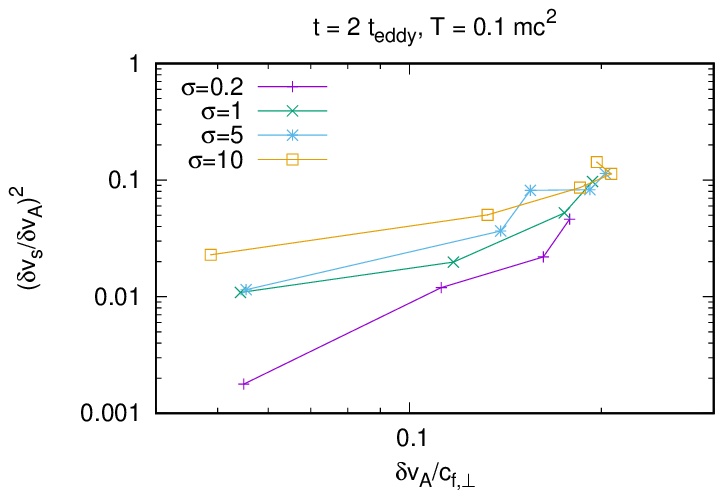}
  \includegraphics[width=8.cm,clip]{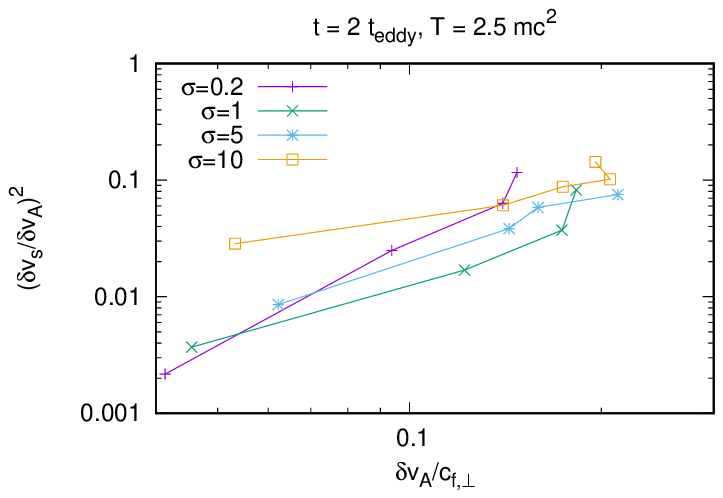}
  \caption{The ratio of slow to Alfv\'en mode power 
           in terms of the non-relativistic fast Mach number 
           %\textcolor{red}{%
           which is measured using the Alfv\'en mode velocity 
           at two eddy-turnover times. 
           %}%
           Top:$T=0.1 mc^2$, Bottom:$T = 2.5 mc^2$.
           }
  \label{fig:1.2}
\end{figure}

Figure \ref{fig:1.2} is the slow to Alfv\'en mode ratio at two eddy-turnover times. 
%where the eddy-turnover time is defined as $t_{\rm eddy} \equiv L/v_{\rm inj}$. 
The top panel is the results of cold plasma case with $T / mc^2 = 0.1$, 
and bottom panel is the results of hot plasma case with $T / mc^2 = 2.5$. 
Although it is difficult to find a clear dependence on the background $\sigma$ parameter 
comparing with the fast to Alfv\'en mode power ratio, 
they also indicate that the ratio slightly increases with the Alfv\'en mode velocity
\footnote{
%\textcolor{red}{%
Note that both Figures \ref{fig:1.1} and \ref{fig:1.2} show decrease of Alfv\'en mode velocity in the case of $\sigma=10$. 
This is because the trans-Alfv\'enic injection velocity, $\lesssim 0.6 c_{\rm A}$, induced a stronger non-linear effects, 
such as formation of shocks. 
This would convert more Alfv\'en mode energy into compressible modes one, 
resulting in reducing Alfv\'en mode velocity with increasing the compressible mode velocity. 
%}%
}
.

\begin{figure}%[t]
 \centering
  \includegraphics[width=8.cm,clip]{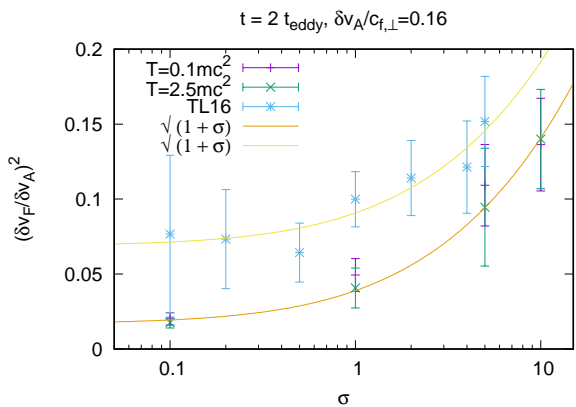}
  \caption{%$f(\sigma) = 0.051 \sqrt{1+\sigma} - 0.0267$, $g(\sigma) = 0.053 \sqrt{1+\sigma} - 0.0367$.
            $\sigma$-dependence of fast to Alfv\'en mode power ratio. 
            TL16 means the adiabatic case studied in TL16. 
            }
  \label{fig:1.3}
\end{figure}

Figure \ref{fig:1.3} is a plot of the fast to Alfv\'en power ratio 
at $\delta v_{\rm A}/c_{\rm f,\perp} = 0.16$ in terms of the background $\sigma$ value. 
The purple and green points are the case of $T=0.1mc^2$ and $T=2.5mc^2$, respectively. 
They indicate that 
both cases can be fitted by a line: 
\begin{equation}
  (\delta v_{\rm F}/\delta v_{\rm A})^2 \simeq A \sqrt{1 + \sigma} (\delta v_{\rm A}/ c_{\rm f,\perp}),
  \label{eq:4.1.1}  
\end{equation}
where $A$ is a parameter independent of background temperature indicated in Figure \ref{fig:1.3}. 
The blue points are in the adiabatic gas case studied in our previous work TL16. 
It shows that both adiabatic and isothermal plasma turbulence show the same dependence on the $\sigma$ value, 
proportional to $\sqrt{1 + \sigma}$
\footnote{
%\textcolor{red}{%
Note that 
Figure \ref{fig:1.3} also plots error bars. 
These result from the fluctuation of each $\sigma$-value's curve in Figure \ref{fig:1.1} 
and \ref{fig:1.2} 
which reflects the statistical fluctuations of initial velocity field. 
This shows that the effect of fluctuation of initial condition increases with the background 
$\sigma$-parameter, 
describing a stronger generation of the compressible modes by the pressure gradient 
force as described in Equation (\ref{eq:4.2.4}), 
resulting in strong temporal fluctuations. 
%}%
}
. 
However, in the adiabatic case the ratio of fast to Alfv\'en power becomes clearly larger than that in the isothermal cases. 
We consider that this is because 
%\textcolor{red}{%
the present prescription of the cooling process 
%}%
breaks the conservation of the total energy, 
and some amount of the initial Alfv\'en mode energy was not converted into fast mode but disappeared from the system through the cooling. 
It also indicates that 
the physics of the non-linear coupling responsible for the mode conversion does not depend on 
%the adiabatic constant of background plasma, 
%\textcolor{red}{%
whether the background plasma is adiabatic or not. 
%}%

%\subsection{\textcolor{red}{A Theoretical Consideration}}
\subsection{A Theoretical Consideration}
\label{sec:sec4.1.1}

In this section, 
we discuss a theoretical consideration of generation of compressible modes by Alfv\'en mode 
in high-$\sigma$ plasma using quasi-linear treatment. 
%Such behaviour can be understood as follows. 
In the non-relativistic case, 
the displacement vector of each MHD mode velocity in low-$\beta$ plasma completely decouples as \citep{2002PhRvL..88x5001C}:
\begin{equation}
  \delta {\bf v}_{\rm A} \propto {\bf e}_{\rm z}, \quad  \delta {\bf v}_{\rm F} \propto {\bf e}_{\rm y}, \quad\delta {\bf v}_{\rm S} \propto {\bf e}_{\rm x}, 
  \label{eq:4.2.1}
\end{equation}
where the same coordinate is selected as in Section \ref{sec:sec3}. 
The magnetic field of these modes can be written as: 
\begin{equation}
  \delta {\bf B}_{\rm A} = - \frac{B_0 \delta v_{\rm A}}{c_{\rm A}} {\bf e}_{\rm z}
  , \quad
  \delta {\bf B}_{\rm F} = \frac{B_0 \delta v_{\rm F}}{c_{\rm A}} [- k_{\rm x} {\bf e}_{\rm y} + k_{\rm y} {\bf e}_{\rm x}]
  , \quad
  \delta {\bf B}_{\rm S} = 0
  .
  \label{eq:4.2.2}
\end{equation}
This shows that magnetic field of each mode also completely decouples. 
%Although these modes still couple in the 2nd-order level in MHD equation, 
%the above behavior makes their coupling minimum. 
In the non-relativistic MHD case, 
the 2nd-order coupling was obtained assuming so called \textit{weak turbulence approximation} 
\citep{2001JETP...93.1052K,2005PhRvL..95z5004C}. 
It is found that 
the interaction between Alfv\'en and fast modes exists through 3-wave resonant interactions. 
In particular, 
it becomes strong at small $\theta$ because the frequency of fast and Alfv\'en modes are comparable, 
which makes the interaction more efficient. 
This interaction tries to make the amount of fast mode comparable to Alfv\'en mode. 
%\textcolor{blue}{%
The faster interaction of pure three-Aflv\'en mode and pure three-fast mode than the mixed interaction, 
%}%
however, prohibits the appearance 
of such a strong-coupling regime~\citep{2005PhRvL..95z5004C}. 
On the other hand, Equation (\ref{eq:3.26}) shows that 
the relativistic correction term breaks such a complete decoupling of the velocity. 
We consider that this results in a stronger coupling. 

The tendency of the fast to Alfv\'en mode velocity ratio increasing with $\sigma$-value in Figure \ref{fig:1.1} can be 
%\textcolor{red}{understood} as follows. 
understood as follows. 
Assuming that only Alfv\'en mode exists in a plasma, 
it induces compressible modes because of its electromagnetic field pressure. 
The equation of motion along background magnetic field can be written as: 
\begin{equation}
  \partial_t [\rho h \delta v_{||}] \simeq - \nabla (\delta B^2 + \delta E^2)/2
  .
  \label{eq:4.2.3}  
\end{equation}
Using the eigen-mode relation of the Alfv\'en mode, $(\delta E_{\rm A})^2 = (\delta v_{\rm A} B_0)^2$ and $(\delta B/B_0)^2 = (\delta v_{\rm A}/c_{\rm A})^2$, 
this reduces to 
\begin{equation}
  \left( \frac{\delta v_{||}}{\delta v_{\rm A}} \right) %\sim \sigma \left( \frac{1 + c_{\rm A}^2}{c_{\rm A}^2} \right) \left( \frac{\delta v_{\rm A}}{c_{\rm A}} \right)
  \sim (\sigma + 1/2)  \left( \frac{\delta v_{\rm A}}{c_{\rm A}} \right)
  ,
  \label{eq:4.2.4}  
\end{equation}
where $\omega = c_{\rm A} k$ and $B_0^2 / \rho h \simeq \sigma$ are used. 
This indicates that the higher the $\sigma$-value is, the more compressible modes are induced; 
On the other hand, 
it becomes independent of $\sigma$-value in the matter-energy dominated region, $\sigma = 0$, 
as indicated in \citep{2002PhRvL..88x5001C}. 
The actual relation of fast and Alfv\'en modes also depends on the energy distribution between fast and slow mode, 
and on non-linear coupling of each modes, 
which will alter the $\sigma$ dependence as indicated in Equation (\ref{eq:4.1.1}). 

%\subsection{\label{sec:sec4.2}Mode Conversion From Each Mode}
\subsection{Energy Transfer Between Modes}
\label{sec:sec4.2}

Since in this work strong turbulence is considered, 
it is very difficult to explain the obtained numerical result thoroughly via an analytical method. 
Hence, in this section a series of numerical experiments were performed, 
and we discuss the mode conversion of each MHD mode in a Poynting-dominated plasma 
with $\sigma=5$. 

\begin{figure}%[t]
 \centering
  \includegraphics[width=8.cm,clip]{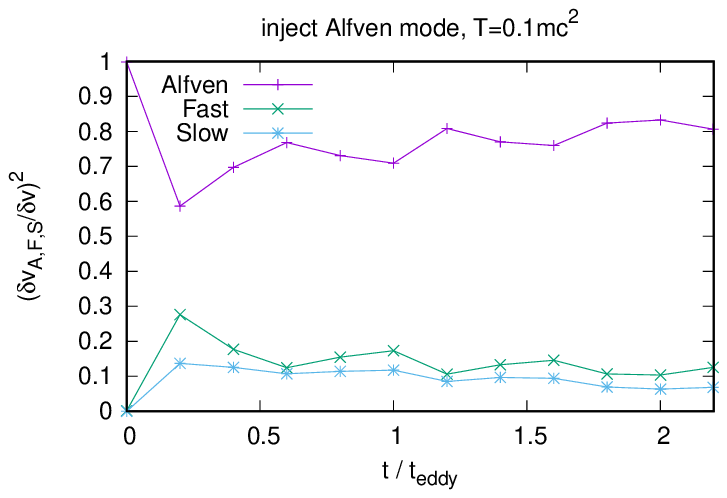}
  \includegraphics[width=8.cm,clip]{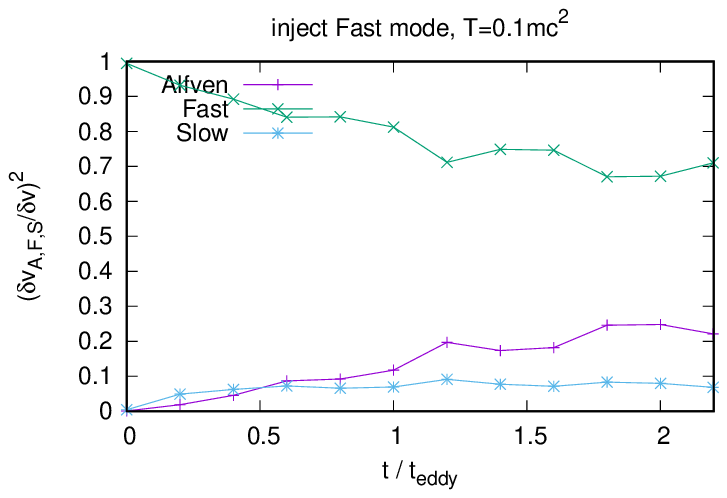}
  \includegraphics[width=8.cm,clip]{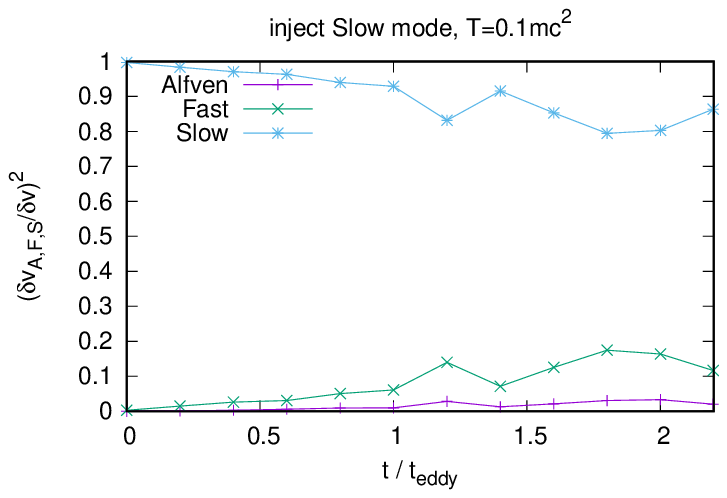}
  \caption{Temporal Evolution of each modes. 
           in the case of $\sigma = 5$ and $T=0.1mc^2$. 
           Top: only Alfv\'en mode is injected. 
           Middle: only fast mode is injected. 
           Bottom: only slow mode is injected. 
           }
  \label{fig:2.1}
\end{figure}

\begin{figure}%[t]
 \centering
  \includegraphics[width=8.cm,clip]{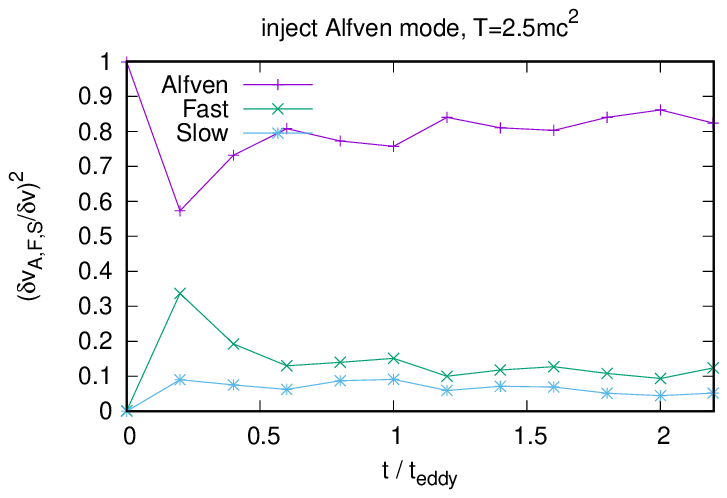}
  \includegraphics[width=8.cm,clip]{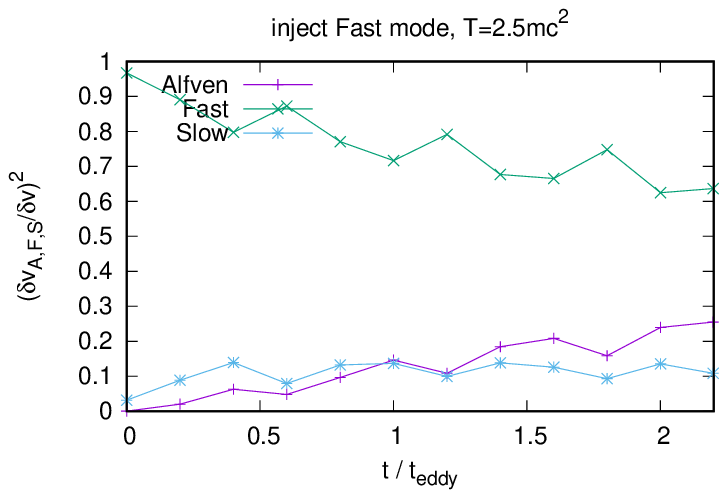}
  \includegraphics[width=8.cm,clip]{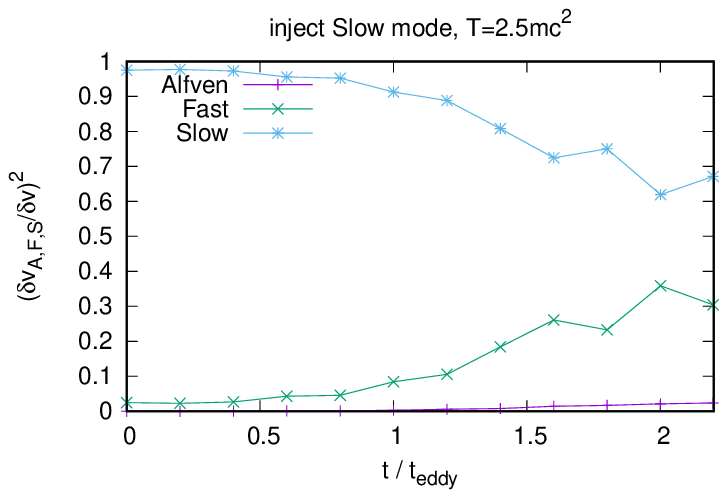}
  \caption{Temporal Evolution of each modes. 
           in the case of $\sigma = 5$ and $T=2.5 mc^2$. 
           Top: only Alfv\'en mode is injected. 
           Middle: only fast mode is injected. 
           Bottom: only slow mode is injected. 
           }
  \label{fig:2.2}
\end{figure}

Figures \ref{fig:2.1} and \ref{fig:2.2} are temporal evolution of each mode velocity 
in the case of $T = 0.1 mc^2$ and $T=2.5 mc^2$, respectively. 
%\textcolor{red}{%
The total injected turbulent velocity dispersion is $\delta v_{\rm inj} = 0.45 c_{\rm A}$. 
%}%
The top panels are the results when injecting only Alfv\'en mode turbulence. 
They show that the mode conversion from Alfv\'en mode to compressible modes is nearly independent of the background temperature, 
consistent with Figure \ref{fig:1.3}. 
In both cases, the each mode energy density in the steady state is approximately described as: $W_{\rm A}:W_{\rm F}:W_{\rm S} \simeq 80:15:5$ 
in the $\sigma = 5$ and isothermal plasma
\footnote{
%\textcolor{red}{%
Note that in the vertical axes in Figures \ref{fig:2.1} and \ref{fig:2.2} the each 
mode power is renormalized by their total power $\delta v^2$. This is the reason 
why the Alfv\'en mode looks like increasing after their sharp decrease at the initial 
phases. The Alfv\'en mode energy itself gradually decreases in time. 
%}%
}
. 

%\begin{figure}[t]
% \centering
%  \includegraphics[width=8.cm,clip]{./graphics/tev_k3v45_sgm5T01_F.eps}
%  \includegraphics[width=8.cm,clip]{./graphics/tev_k3v45_sgm5T25_F.eps}
%  \caption{Temporal Evolution of each modes when only fast mode is injected 
%           in the case of $\sigma = 5$. 
%           Left: $T=0.1mc^2$, Right: $T=2.5 mc^2$
%           }
%  \label{fig:2.2}
%\end{figure}

The middle panels of Figures \ref{fig:2.1} and \ref{fig:2.2} are temporal evolution of each mode velocity 
when injecting only fast mode turbulence. 
%The left panel is the case of $T = 0.1 mc^2$, and right panel is the case of $T=2.5 mc^2$. 
They also show that the mode conversion from fast mode to the other modes is nearly independent of the background temperature. 
In both cases, the each mode energy in the steady state is approximately described as: $W_{\rm A}:W_{\rm F}:W_{\rm S} \simeq 20:70:10$, 
indicating a stronger energy transfer between Alfv\'en and fast modes. 

%\begin{figure}[t]
% \centering
%  \includegraphics[width=8.cm,clip]{./graphics/tev_k3v45_sgm5T01_S.eps}
%  \includegraphics[width=8.cm,clip]{./graphics/tev_k3v45_sgm5T25_S.eps}
%  \caption{Temporal Evolution of each modes when only slow mode is injected 
%           in the case of $\sigma = 5$. 
%           Left: $T=0.1mc^2$, Right: $T=2.5 mc^2$
%           }
%  \label{fig:2.3}
%\end{figure}

The bottom panels of Figures \ref{fig:2.1} and  \ref{fig:2.2} are temporal evolution of each mode velocity 
when injecting only slow mode turbulence. 
%The left panel is the case of $T = 0.1 mc^2$, and right panel is the case of $T=2.5 mc^2$. 
In contrast to the previous cases, 
they show a clear dependence of the slow to fast energy conversion on the background temperature. 
We consider that 
this may indicate the effect of the relativistic correction term in Equation (\ref{eq:3.26}) 
which induces a mixing of the velocity direction between fast and slow mode. 
Although it is very difficult to explain why slow to fast mode conversion seems much stronger than the opposite case 
due to the strong non-linearity in the turbulence, 
at least we found even in Figures \ref{fig:2.1} and \ref{fig:2.2} a slight increase of the mode conversion into slow mode can be observed. 
In Figure \ref{fig:2.1}, the each mode energy in the steady state is approximately described as: $W_{\rm A}:W_{\rm F}:W_{\rm S} \simeq 0:10:90$, 
and in Figure \ref{fig:2.2}, $W_{\rm A}:W_{\rm F}:W_{\rm S} \simeq 0:30:70$. 

%\textcolor{red}{%
Note that Figures \ref{fig:2.1} and \ref{fig:2.2} show that 
the initial Alfv\'en mode power is immediately transferred to the other modes, mainly
into fast mode, much less than the eddy-turnover time 
which is the typical timescale of energy transfer in the case of the 
critical-balanced Alfv\'en mode turbulence. 
This is because in this regime the initial energy is not transferred by the 
mode-coupling but the pressure gradient force. 
The timescale of the energy conversion via the pressure gradient force can be
estimated using Equation (\ref{eq:4.2.3}) as follows. 
In the case of the initial energy transfer, Equation (\ref{eq:4.2.3}) can be rewritten 
as:
\begin{equation}
  \frac{\rho h \delta v_{||}}{\tau_{\rm trans}} \sim \frac{\delta B^2 + \delta E^2}{2 L_{\rm inj}}, 
  \label{eq:4.2.0.1}
\end{equation}
where $L_{\rm inj}$ is the initial injection scale, and $\tau_{\rm trans}$ is the energy transfer timescale. 
Assuming $\delta v_{||} \sim r_0 \delta v_{\rm A,0}, \delta B / B_0 \sim \delta v_{\rm A,0}/c_{\rm A}, \delta E \sim \delta v_{\rm A,0} B_0$, 
the above equation reduces to: 
\begin{equation}
  \tau_{\rm trans} \sim L_{\rm inj} \ \frac{r_0}{\delta v_{\rm A,0}^2} \ \frac{2}{2 \sigma + 1}
  .
\end{equation}
Note that $\tau_{\rm trans}$ describes the energy transfer timescale of the evolution of $\delta v_{||}$ from 0 to $r_0 \delta v_{\rm A,0}$ 
where $\delta v_{\rm A,0}$ is the initial Alfv\'en mode velocity and $r_0$ is a constant coefficient. 
Substituting $L_{\rm inj} \sim L/3, r_0 \sim 1$, and $\sigma = 5$, 
the timescale becomes $\tau_{\rm trans}/(L/\delta v_{\rm A,0}) \sim 0.06$, 
which explains the initial rapid energy transfer described in Figures \ref{fig:2.1}, \ref{fig:2.2}, and \ref{fig:2.4}. 
%}%

\subsection{A Strong Coupling Regime}
\label{sec:sec4.3}

\begin{figure}%[t]
 \centering
  \includegraphics[width=8.cm,clip]{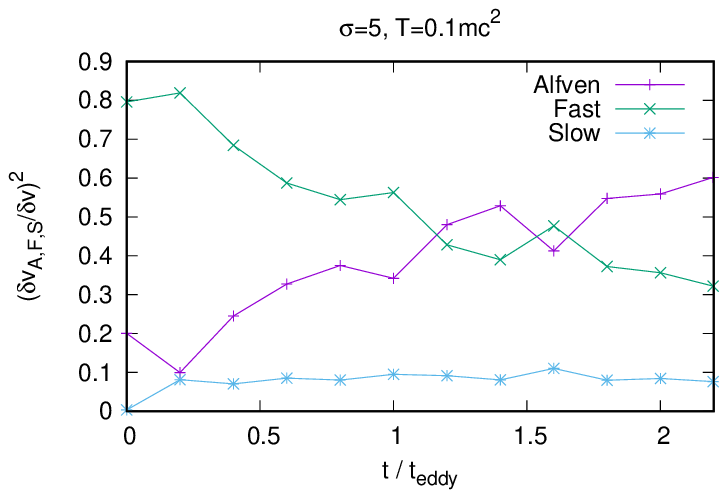}
  \includegraphics[width=8.cm,clip]{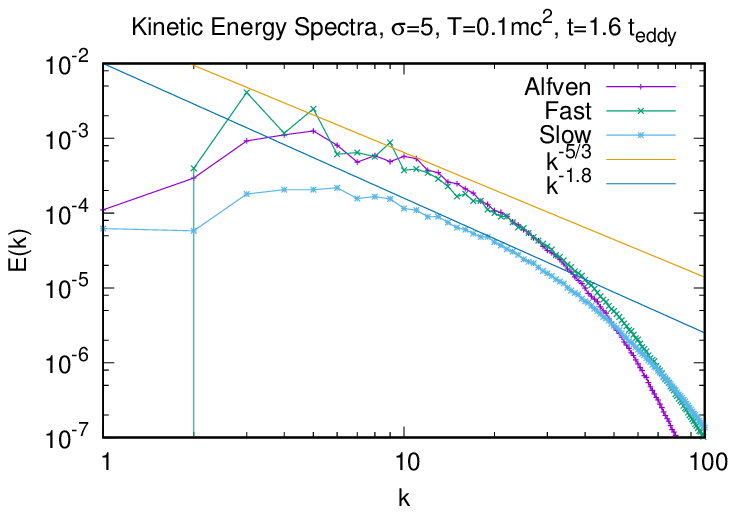}
  \caption{Top:Temporal Evolution of each modes when injecting $(\delta v_A)^2:(\delta v_F)^2 = 1:4$ 
           in the case of $\sigma = 5$ and $T=0.1mc^2$.
           Bottom: kinetic energy spectra of each mode at $t=1.6t_{\rm eddy}$.
           }
  \label{fig:2.4}
\end{figure}

As we have seen in the previous sections, 
the obtained results indicate existing of a region where fast and Alfv\'en modes are strongly coupled, 
different from non-relativistic cases. 
From Figure \ref{fig:1.3}, 
the coefficient A in Equation (\ref{eq:4.1.1}) is 0.33 in the case of isothermal plasma. 
Assuming $\delta v_{\rm A}/c_{\rm f,\perp} = 0.5$, 
the indicated $\sigma$-value necessary for this regime, $\delta v_{\rm F} \simeq \delta v_{\rm A}$, becomes $\sigma = 35$. 
Unfortunately, such a high-$\sigma$ plasma is very difficult to 
simulate by the present numerical simulation technique. 
%Unfortunately, the indicated $\sigma$-value necessary for this regime in Figure \ref{fig:1.3} is very difficult to 
%be performed by the present numerical simulation technique. 
Instead, 
we performed a run with a special initial condition that can temporally results in the strong regime, 
and studied its physical properties. 
The initial injected turbulence was composed by 80 \% fast mode and 20 \% Alfv\'en mode power, 
and the background plasma parameters are: $\sigma = 5$ and $T/mc^2 = 0.1$. 
%\textcolor{red}{%
The total injected turbulent velocity dispersion is $\delta v_{\rm inj} = 0.45 c_{\rm A}$. 
%}%
The other conditions were set the same as the other runs. 
The temporal evolution of each mode power is shown in the top-panel of Figure \ref{fig:2.4}. 
It indicates that 
fast and Alfv\'en modes strongly couple around 1.5 eddy-turnover time. 
After that time, 
the coupling became weak, and the turbulence gradually evolved into a state indicated in Figure \ref{fig:1.3} 
because of insufficient background $\sigma$-value. 
In the following, we concentrate on the strong coupling regime appeared in this simulation. 
The bottom-panel of Figure \ref{fig:2.4} is the energy spectra of each velocity components at t = 1.6 eddy-turnover time. 
First, 
it is found that strong fluctuations exist in the long wavelength region, 
and this makes it difficult to find a clear inertial region. 
In spite of this fact, it shows the energy spectra of fast and Alfv\'en modes clearly degenerates up to dissipation regime around $k=40$; 
This indicates that the strong coupling regime of the 2 mode was actually realized in this simulation. 
Its spectral index up to $k=20$ is approximately shallower that $5/3$ but steeper than $1.8$ 
which is observed in fast mode energy spectrum in our previous work TL16. 
However, 
the value of the index is clearly closer to $5/3$ than $1.8$, 
and this indicates that the energy cascade into smaller scale length is mainly driven by Alfv\'en mode-like cascade process. 

\begin{figure}%[t]
 \centering
  \includegraphics[width=8.cm,clip]{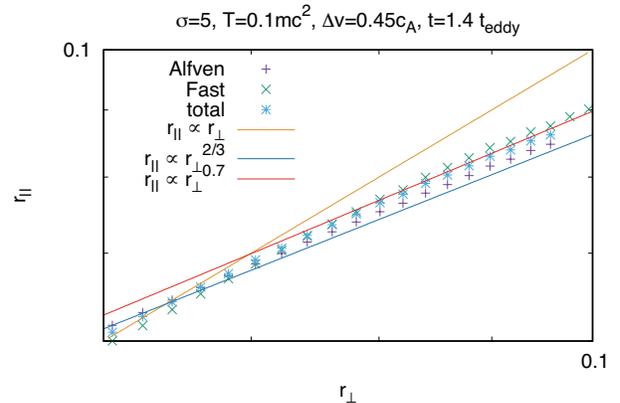}
  \caption{Eddy-shape of Alfv\'en and fast modes in the case of $\sigma=5$ and $T=0.1 mc^2$ 
           at $t=1.4 t_{\rm eddy}$ The integrated case (no mode decomposition) is also plotted, 
           which shows an intermediate region between isotropic: $r_{||} \propto r_{\perp}$ to 
           critical balance case $r_{||} \propto r_{\perp}^{2/3}$. The indicated power law index 
           approximately $0.7$. 
           }
  \label{fig:2.5}
\end{figure}

Figure \ref{fig:2.5} shows the eddy-shape of the Alfv\'en, fast, and the total mode of the 2-modes. 
Following \citep{2000ApJ...539..273C}, 
the eddy-shape is calculated from the 2nd-order structure function for the velocity. 
Note that 
the structure function is calculated in terms of the local magnetic field \citep{1999ApJ...517..700L,2000ApJ...539..273C,2001ApJ...554.1175M}, 
and the axes of the figure $(r_{\perp}, r_{||})$ means the radius of the eddy 
perpendicular and parallel to the local magnetic field. 
The figure shows that 
all the 3-lines can be described as: $r_{||} \propto r_{\perp}^{0.7}$ 
which is close to the critical balance law:  $r_{||} \propto r_{\perp}^{2/3}$. 
This also supports that 
the energy cascade is mainly governed by the Alfv\'en mode-like. 
This can be because 
the fast mode cascade is less efficient than the Alfv\'en mode cascade, 
and the strong coupling allows the fast mode energy to flow into Alfv\'en mode 
whose cascade is more efficient. 
%which is indicated by a softer energy spectrum of fast mode $E(k) \propto k^{-1.86}$ 
%as shown in TL16. 
The power law index is a little larger than the critical value $2/3$, 
and we consider that 
this correction is due to an effect of fast mode 
which usually shows isotropic eddy-shape $r_{||} \propto r_{\perp}$ 
as shown in TL16. 
Although the obtained results are just a state of temporally strong-coupling, 
we expect that it indicates the existence of the steady strong-coupling regime 
where a similar properties will be able to be observed. 

\section{Discussion}
\label{sec:sec5}
%\clearpage
\begin{figure}%[t]
 \centering
  \includegraphics[width=8.cm,clip]{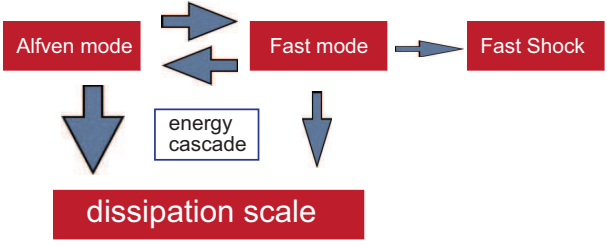}
  \caption{A schematic picture of the energy transfer in high-$\sigma$ turbulence. 
           }
  \label{fig:2.6}
\end{figure}

Although it was impossible to show an existence of the strong-coupling regime due to the numerical difficulty 
of simulating strong turbulence in high-$\sigma$ plasma, 
we simulated a temporally strongly-coupling regime. 
In that regime, 
it is found that 
the energy spectra of Alfv\'en and fast modes are nearly the same 
whose spectral index is nearly 5/3, 
possibly indicating the energy cascade is mainly via Alfv\'en mode-like cascade. 
We also studied the eddy shape in the strong-coupling regime. 
The obtained eddy shape is approximately $r_{||} \propto r_{\perp}^{0.7}$ 
which is slightly different from the shape indicated by the critical-balance, $r_{||} \propto r_{\perp}^{2/3}$. 
We consider that 
this is due to the effect of the coupling to fast mode 
whose eddy shape is isotropic, $r_{||} \propto r_{\perp}$. 
Although the obtained results are just a state of temporally strong-coupling, 
we expect that it indicates the existence of the steady strong-coupling regime 
where a similar properties will be able to be observed. 

%\textcolor{red}{%
From these facts, 
one possible scenario can be proposed for the turbulence-energy transfer in the strong-coupling regime. 
The degeneration of the fast and Alfv\'en modes in Figures \ref{fig:2.5} and \ref{fig:2.6} indicates that 
the Alfv\'en and fast modes communicate with each other in large scale and inertial region, 
allowing energy transfer between the 2-modes; 
This coupling, however, cannot be observed in the small scale region, or the dissipation region in Figure \ref{fig:2.5}. 
This means that the energy dissipation process of Alfv\'en and fast modes are different. 
We consider that a possible candidate for the dissipation process of fast mode is the appearance of the fast-shock. 
It is well-known that the fast wave evolves into fast shock due to the non-linearity, 
and this allows to dissipate its own energy through shock dissipation as discussed in \citep{2014ApJ...787...84T}. 
Concerning the inertial regime, 
the eddy-shape and the spectral index of the energy spectrum indicate that 
the energy will be transferred mainly by the Alfv\'en mode-like cascade process, 
and a small amount of it by the fast mode-like cascade. 
In summary, 
it can be expected that 
there will be an energy cascade path of turbulence in high-$\sigma$ plasma 
as shown in Figure \ref{fig:2.6}. 
%}%
\section{Summary and Conclusion}
\label{sec:sec6}

%\textcolor{red}{%
In this paper, 
we discussed a strong-coupling regime of Alfv\'en and fast modes of relativistic MHD turbulence in Poynting-dominated plasma. 
In our previous work TL16, 
we found an increase of the amount of the fast mode with background $\sigma$-value in the case of decaying turbulence in adiabatic plasma, 
which we consider as an indirect proof of strong-coupling between Alfv\'en and fast mode in sufficiently high-$\sigma$ plasma. 
It was, however, not clear if the increase of the amount of fast mode was affected by the increase of background temperature 
which reduces $\sigma$-value and may change the behavior of the mode exchange of RMHD turbulence. 
For this problem, 
in this work, isothermal plasma was considered that kept background plasma temperature constant. 
We found that 
the increase of the fast to Alfv\'en mode power ratio can be observed even in this case as shown in Figure \ref{fig:1.1}. 
We consider that 
this is because the relativistic correction of the displacement vector of fast mode in high-$\sigma$ plasma 
which will encourage mixing of fast to the other two modes through 2nd-order interaction. 
We also performed a series of numerical studies of mode conversion. 
It showed that 
Alfv\'en mode turbulence actually generate fast mode turbulence in high-$\sigma$ plasma, and vice versa.
In addition, 
we found that 
the mode exchange is nearly insensitive to the background temperature, 
other than the slow to fast mode conversion. 
%}% 

In conclusion, 
we found that 
the strong-coupling regime of Alfv\'en to fast modes was independent of the background temperature, 
and actually existed in high-$\sigma$ plasma turbulence. 
We emphasize that 
this alters the previous understanding on relativistic turbulence in high-energy astrophysical phenomena, 
which usually assumed either the non-relativistic critical balance or force-free limit. 
In this sense, 
our new findings will affect the understanding of particle acceleration and photon emission in high-$\sigma$ plasma turbulence, 
such as relativistic jets, pulsar wind nebulae, and gamma-ray bursts. 

\section*{Acknowledgments}
We would like to thank S\'ebastien Galtier, Supratik Banerjee, and Nobumitsu Yokoi 
for many fruitful comments and discussions. 
MT also would like to thank Tsuyoshi Inoue for kindly providing with the original numerical code. 
Numerical computations were carried out on the Cray XC30 
at Center for Computational Astrophysics, CfCA, of National Astronomical Observatory of Japan.
%Calculations were also carried out on SR16000 at YITP in Kyoto University. 
This work is supported in part by the Postdoctoral Fellowships by the Japan Society for the Promotion of Science No. 201506571 (M. T.). 
%AL is supported by NSF AST 1212096, NASA grant X5166204101. 
%\textcolor{red}{%
AL is supported by NSF DMS 1622353. 
%}%
% and of the NSF sponsored Center for Magnetic Self-Organization. 

%I thank Professor N. Kameswara Rao for some helpful suggestions,
%Dr H. C. Bhatt for a critical reading of the original version of the
%paper and an anonymous referee for very useful comments that improved
%the presentation of the paper.

\appendix

\section{Energy Spectra of Driven Turbulence in Isothermal Plasmas}
\label{sec:secA1}

\begin{figure}%[t]
 \centering
  \includegraphics[width=8cm,clip]{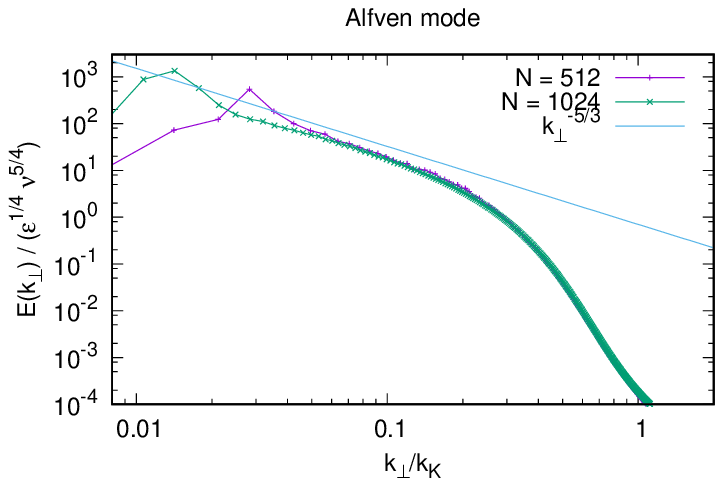}
  \includegraphics[width=8cm,clip]{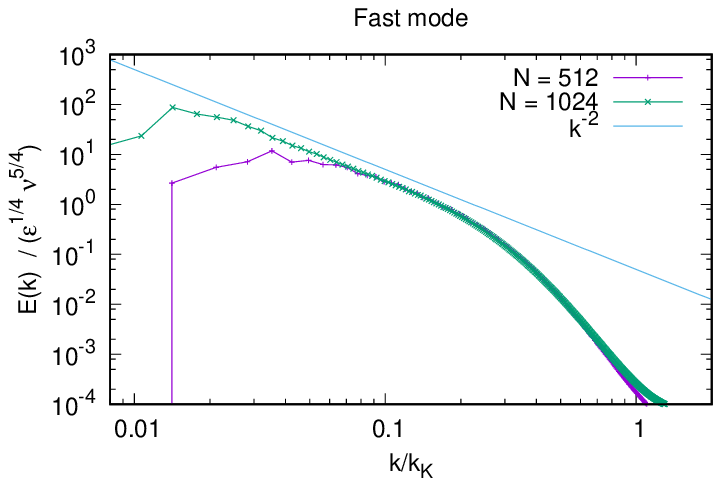}
  \includegraphics[width=8cm,clip]{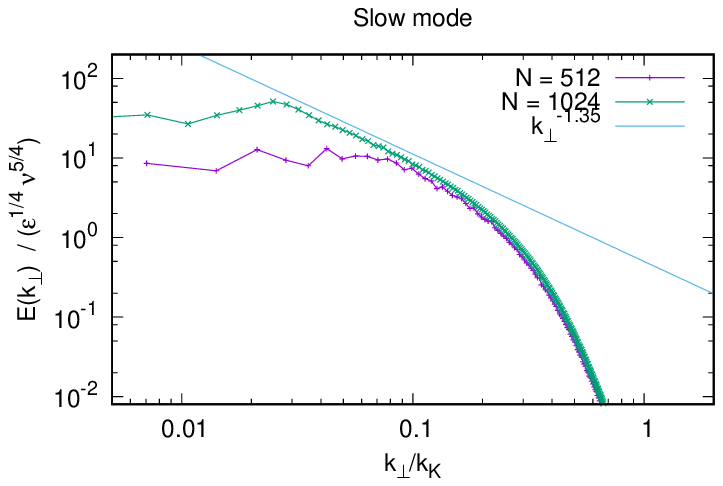}
  \caption{Non-dimensional energy spectra of each mode in driven turbulence in isothermal plasma 
           with $\sigma=3$ and $T/mc^2 = 0.1$. 
           }
  \label{fig:A1}
\end{figure}

In this Appendix, 
we discuss energy spectra of each mode in high-$\sigma$ turbulence. 
In our previous work TL16, 
the numerical simulations of undriven turbulence in adiabatic plasma were performed, 
and the obtained energy spectra were discussed. 
On the other hand, 
in this work, so called \textit{driven} turbulence is considered, 
which allows us to obtain turbulence without suffering from contamination by dissipation in its energy spectra. 
Note that we also consider an effect of cooling to avoid too much energy increase by injecting turbulence. 

In this work, 
turbulence is injected by adding to momentum flux density, $\rho h \gamma^2 \delta {\bf v}$, 
where $\delta {\bf v}$ is a turbulent velocity obtained the same method as described in Section \ref{sec:sec2}. 
The background plasma parameters are $\sigma = 3$ and $T / m c^2 = 0.1$. 
The strength of the injected turbulence is $|\delta v| = 0.5 c_{\rm A}$. 
The turbulence is injected at every fixed timestep, $dt_{\rm inj} = \sqrt{2} t_{\rm eddy}/10$
\footnote{
We saved data at every $0.2 t_{\rm eddy}$. 
The factor $\sqrt{2}$ is added to avoid injecting turbulence at the same time step of data saving. 
}. 

The main purpose of this section is to discuss properties of inertial region of each mode. 
Unfortunately, it is usually very difficult to find clear inertial region of relativistic MHD turbulence 
because of insufficient numerical resolutions. 
To avoid such a problem, 
a new strategy is proposed by \citep{2014ApJ...784L..20B,2015ASSL..407..163B}, 
which uses the self-similar property of turbulence, 
so-called \textit{Kolmogorov's similarity law} \citep{1941DoSSR..30..301K},
originally formulated in the case of pure hydrodynamic turbulence. 
This theory predicts that 
the energy spectrum of scale-invariant turbulence should be described as: 
\begin{equation}
  E(k) = \nu^{5/4} \epsilon^{1/4} e(k/k_{\rm K}), 
\end{equation}
where $\nu$ is kinetic viscosity, 
$\epsilon$ is the energy cascade rate, 
$e(k)$ is a non-dimensional power-law distribution function, 
and 
\begin{equation}
  k_{\rm K} \equiv \left(\frac{\epsilon}{\nu} \right)^{1/4}
  ,
\end{equation}
is the Kolmogorov wavenumber describing a characteristic length of dissipation. 
This tells that 
the non-dimensional energy spectrum, $e(k/k_{\rm K}) = E(k/k_{\rm K})/(\nu^{5/4} \epsilon^{1/4})$, should be independent of the background Reynolds number, 
and a clear inertial region can be found by comparing the simulation results with different resolution or Reynolds number. 

Figure \ref{fig:A1} is the obtained non-dimensional energy spectra of each mode 
at $t = 2 t_{\rm eddy}$ when the turbulence reached a steady state
\footnote{
The energy spectra were obtained using a framework for parallel computations of Fourier transforms in three dimensions \citep{doi:10.1137/11082748X}. 
}
.
We assumed that 
$\nu$ can be written as $\nu = C c_{\rm s} \Delta x$ 
where $\Delta x$ is the mesh size, 
and $C$ is a constant value describing numerical dissipation in our simulation
\footnote{For the value of $C$, we took unity when $N=512$ and 0.8 when $N=1024$}. 
They indicate that 
the relativistic MHD turbulence in high-$\sigma$ plasma follows the Kolmogorov's similarity law. 
The obtained energy spectra in the inertial region can be read approximately as $-5/3, -2, -1.35$ in the case of Alfv\'en, fast, and slow modes, respectively. 
The value of the index of Alfv\'en mode, -5/3, is consistent with our previous work TL16. 
Interestingly, 
the fast mode becomes much steeper than that of decaying turbulence, -1.86, obtained in TL16. 
On the other hand, 
the slow mode shows a very shallow spectrum, -1.35, 
although TL16 cannot have found a clear inertial region of slow mode turbulence. 
It is still unclear that 
the observed spectra suffered the injection scale or not, 
though it seems not affected from dissipation scale. 
And it is necessary to perform much larger simulation in the future in order to obtain a robust conclusion. 

%\textcolor{blue}{%
Finally, 
in this simulation, sub-Alfv\'enic turbulence is injected, 
and this induces a mixed regime of weak and strong turbulence in long and short wave length regime \citep{1999ApJ...517..700L}. 
In Figure \ref{fig:A1}, however, we cannot find a weak regime characterized by the spectral index $-2$. 
We consider that 
this is due to the resolution used in this simulation. 
The transition from weak to strong turbulence was first observed in \citep{2016PhRvL.116j5002M}, 
which used more than 3000 meshed in the direction perpendicular to the background magnetic field. 
In addition, 
we considered a magnetically-dominated plasma, 
which demands, in general, higher resolution to resolve the energy cascade in the perpendicular direction. 
%}%

%\bibliography{RCT}% Produces the bibliography via BibTeX.

\begin{thebibliography}{}

\bibitem[\protect\citeauthoryear{{Anile}}{{Anile}}{1990}]{1990rfmf.book.....A}
{Anile} A.~M.,  1990, {Relativistic Fluids and Magneto-fluids}

\bibitem[\protect\citeauthoryear{{Ant{\'o}n}, {Miralles}, {Mart{\'{\i}}},
  {Ib{\'a}{\~n}ez}, {Aloy} \& {Mimica}}{{Ant{\'o}n}
  et~al.}{2010}]{2010ApJS..188....1A}
{Ant{\'o}n} L.,  {Miralles} J.~A.,  {Mart{\'{\i}}} J.~M.,  {Ib{\'a}{\~n}ez}
  J.~M.,  {Aloy} M.~A.,    {Mimica} P.,  2010, ApJS, 188, 1

\bibitem[\protect\citeauthoryear{{Balsara}}{{Balsara}}{2001}]{2001ApJS..132...83B}
{Balsara} D.,  2001, ApJS, 132, 83

\bibitem[\protect\citeauthoryear{{Beresnyak}}{{Beresnyak}}{2014}]{2014ApJ...784L..20B}
{Beresnyak} A.,  2014, ApJL, 784, L20

\bibitem[\protect\citeauthoryear{{Beresnyak} \& {Lazarian}}{{Beresnyak} \&
  {Lazarian}}{2015}]{2015ASSL..407..163B}
{Beresnyak} A.,  {Lazarian} A.,  2015, in {Lazarian} A.,  {de Gouveia Dal Pino}
  E.~M.,   {Melioli} C.,  eds, Magnetic Fields in Diffuse Media Vol.~407 of
  Astrophysics and Space Science Library, {MHD Turbulence, Turbulent Dynamo and
  Applications}.
p.~163

\bibitem[\protect\citeauthoryear{{Chandran}}{{Chandran}}{2005}]{2005PhRvL..95z5004C}
{Chandran} B.~D.~G.,  2005, Physical Review Letters, 95, 265004

\bibitem[\protect\citeauthoryear{{Cho}}{{Cho}}{2005}]{2005ApJ...621..324C}
{Cho} J.,  2005, ApJ, 621, 324

\bibitem[\protect\citeauthoryear{{Cho} \& {Lazarian}}{{Cho} \&
  {Lazarian}}{2002}]{2002PhRvL..88x5001C}
{Cho} J.,  {Lazarian} A.,  2002, Physical Review Letters, 88, 245001

\bibitem[\protect\citeauthoryear{{Cho} \& {Lazarian}}{{Cho} \&
  {Lazarian}}{2003}]{2003MNRAS.345..325C}
{Cho} J.,  {Lazarian} A.,  2003, MNRAS, 345, 325

\bibitem[\protect\citeauthoryear{{Cho} \& {Lazarian}}{{Cho} \&
  {Lazarian}}{2014}]{2014ApJ...780...30C}
{Cho} J.,  {Lazarian} A.,  2014, ApJ, 780, 30

\bibitem[\protect\citeauthoryear{{Cho}, {Lazarian} \& {Vishniac}}{{Cho}
  et~al.}{2002}]{2002ApJ...564..291C}
{Cho} J.,  {Lazarian} A.,    {Vishniac} E.~T.,  2002, ApJ, 564, 291

\bibitem[\protect\citeauthoryear{{Cho} \& {Vishniac}}{{Cho} \&
  {Vishniac}}{2000}]{2000ApJ...539..273C}
{Cho} J.,  {Vishniac} E.~T.,  2000, ApJ, 539, 273

\bibitem[\protect\citeauthoryear{{Evans} \& {Hawley}}{{Evans} \&
  {Hawley}}{1988}]{1988ApJ...332..659E}
{Evans} C.~R.,  {Hawley} J.~F.,  1988, ApJ, 332, 659

\bibitem[\protect\citeauthoryear{{Gardiner} \& {Stone}}{{Gardiner} \&
  {Stone}}{2005}]{2005JCoPh.205..509G}
{Gardiner} T.~A.,  {Stone} J.~M.,  2005, Journal of Computational Physics, 205,
  509

\bibitem[\protect\citeauthoryear{{Goldreich} \& {Sridhar}}{{Goldreich} \&
  {Sridhar}}{1995}]{1995ApJ...438..763G}
{Goldreich} P.,  {Sridhar} S.,  1995, ApJ, 438, 763

\bibitem[\protect\citeauthoryear{{Inoue}, {Asano} \& {Ioka}}{{Inoue}
  et~al.}{2011}]{2011ApJ...734...77I}
{Inoue} T.,  {Asano} K.,    {Ioka} K.,  2011, ApJ, 734, 77

\bibitem[\protect\citeauthoryear{{Kolmogorov}}{{Kolmogorov}}{1941}]{1941DoSSR..30..301K}
{Kolmogorov} A.,  1941, Akademiia Nauk SSSR Doklady, 30, 301

\bibitem[\protect\citeauthoryear{{Komissarov}}{{Komissarov}}{1999}]{1999MNRAS.303..343K}
{Komissarov} S.~S.,  1999, MNRAS, 303, 343

\bibitem[\protect\citeauthoryear{{Kuznetsov}}{{Kuznetsov}}{2001}]{2001JETP...93.1052K}
{Kuznetsov} E.~A.,  2001, Soviet Journal of Experimental and Theoretical
  Physics, 93, 1052

\bibitem[\protect\citeauthoryear{{Lazarian} \& {Vishniac}}{{Lazarian} \&
  {Vishniac}}{1999}]{1999ApJ...517..700L}
{Lazarian} A.,  {Vishniac} E.~T.,  1999, ApJ, 517, 700

\bibitem[\protect\citeauthoryear{{Maron} \& {Goldreich}}{{Maron} \&
  {Goldreich}}{2001}]{2001ApJ...554.1175M}
{Maron} J.,  {Goldreich} P.,  2001, ApJ, 554, 1175

\bibitem[\protect\citeauthoryear{{Meyrand}, {Galtier} \& {Kiyani}}{{Meyrand}
  et~al.}{2016}]{2016PhRvL.116j5002M}
{Meyrand} R.,  {Galtier} S.,    {Kiyani} K.~H.,  2016, Physical Review Letters,
  116, 105002

\bibitem[\protect\citeauthoryear{{Mignone}, {Plewa} \& {Bodo}}{{Mignone}
  et~al.}{2005}]{2005ApJS..160..199M}
{Mignone} A.,  {Plewa} T.,    {Bodo} G.,  2005, ApJS, 160, 199

\bibitem[\protect\citeauthoryear{{Mignone}, {Ugliano} \& {Bodo}}{{Mignone}
  et~al.}{2009}]{2009MNRAS.393.1141M}
{Mignone} A.,  {Ugliano} M.,    {Bodo} G.,  2009, MNRAS, 393, 1141

\bibitem[\protect\citeauthoryear{Pekurovsky}{Pekurovsky}{2012}]{doi:10.1137/11082748X}
Pekurovsky D.,  2012, SIAM Journal on Scientific Computing, 34, C192

\bibitem[\protect\citeauthoryear{{Radice} \& {Rezzolla}}{{Radice} \&
  {Rezzolla}}{2013}]{2013ApJ...766L..10R}
{Radice} D.,  {Rezzolla} L.,  2013, ApJL, 766, L10

\bibitem[\protect\citeauthoryear{Synge}{Synge}{1957}]{synge1957relativistic}
Synge J.~L.,  1957, The relativistic gas.
Vol.~32, North-Holland Amsterdam

\bibitem[\protect\citeauthoryear{{Takamoto}, {Inoue} \& {Lazarian}}{{Takamoto}
  et~al.}{2015}]{2015ApJ...815...16T}
{Takamoto} M.,  {Inoue} T.,    {Lazarian} A.,  2015, ApJ, 815, 16

\bibitem[\protect\citeauthoryear{{Takamoto}, {Kisaka}, {Suzuki} \&
  {Terasawa}}{{Takamoto} et~al.}{2014}]{2014ApJ...787...84T}
{Takamoto} M.,  {Kisaka} S.,  {Suzuki} T.~K.,    {Terasawa} T.,  2014, ApJ,
  787, 84

\bibitem[\protect\citeauthoryear{{Takamoto} \& {Lazarian}}{{Takamoto} \&
  {Lazarian}}{2016}]{2016ApJ...831L..11T}
{Takamoto} M.,  {Lazarian} A.,  2016, ApJL, 831, L11

\bibitem[\protect\citeauthoryear{{Thompson} \& {Blaes}}{{Thompson} \&
  {Blaes}}{1998}]{1998PhRvD..57.3219T}
{Thompson} C.,  {Blaes} O.,  1998, PRD, 57, 3219

\bibitem[\protect\citeauthoryear{{Zrake} \& {MacFadyen}}{{Zrake} \&
  {MacFadyen}}{2012}]{2012ApJ...744...32Z}
{Zrake} J.,  {MacFadyen} A.~I.,  2012, ApJ, 744, 32

\bibitem[\protect\citeauthoryear{{Zrake} \& {MacFadyen}}{{Zrake} \&
  {MacFadyen}}{2013}]{2013ApJ...763L..12Z}
{Zrake} J.,  {MacFadyen} A.~I.,  2013, ApJL, 763, L12

\end{thebibliography}

\bsp

%\label{lastpage}

\end{document}